# DAOs' Business Value from an Open Systems Perspective: A Best-Fit Framework Synthesis

Lukas Küng[1, 2[0000-0001-5924-4542]] and George M. Giaglis[1[0000-0002-7824-633X]]

[1] University of Nicosia, School of Business, Institute For The Future, Cyprus
[2] ZHAW, Zurich University of Applied Sciences, School of Management and Law

**Abstract.** Decentralized autonomous organizations (DAOs) are emerging innovative organizational structures, enabling collective coordination, and reshaping digital collaboration. Despite the promising and transformative characteristics of DAOs, the potential technological advancements and the understanding of the business value that organizations derive from implementing DAO characteristics are limited. This research applies a systematic review of DAOs' business applicability from an open systems perspective following a *best-fit framework* methodology. Within our approach, combining both framework and thematic analysis, we discuss how the open business principles apply to DAOs and present a new *DAO business framework* comprising of four core business elements: i) token, ii) transactions, iii) value system and iv) strategy with their corresponding sub-characteristics. This paper offers a preliminary DAO business framework that enhances the understanding of DAOs' transformative potential and guides organizations in innovating more inclusive business models (BMs), while also providing a theoretical foundation for researchers to build upon.

## 1 Introduction

DAO business characteristics are promising and evolving, yet the understanding of the DAO leading to new sources of decentralized value and BMs is limited and requires extensive investigation and research on the subject (Kimani et al., 2020; Frizzo-Barker et al., 2020; Bellavitis et al., 2023). Additionally, the current gap between the technological potential and actual business value for organizations to apply DAO innovations is large, causing uncertainty and disillusionment (Ostern et al., 2022). While DAOs as new types of organizational structures are gaining momentum and are reshaping digital communities, communication, and collaboration (Wang et al., 2023), DAO implementation literature is rare (Lustenberger et al., 2024), nevertheless, DAO research is important as it is becoming the norm of future decentralized organization design (Saurabh et al., 2023). While Morkunas et al. (2019) initially discussed the applicability of blockchain to the business model canvas, Saurabh et al. (2023) introduced a first conceptual framework on DAOs applying the nine elements of the business model canvas by Osterwalder and Pigneur (2010) to DAOs. While these studies are valuable contributions to the understanding of the blockchain and DAO BMs, the question of the changing value system through the application of DAOs within an organization's context and the transformative potential of DAOs' business value remain unclear. Following the call for more research on DAOs' applicability and scalability in business value discovery and BMs by Santana and Albaraeda (2022), and highlighting the lack of DAO implementation literature by Saurabh et al. (2023), this review will take the perspective from open systems, in trying to understand the subsequent challenge of developing business opportunities that outline the core causal connections between open value creation and value capture in DAOs. Hence, based on the literature, we explore the characteristics of open innovation and open business models, synthesizing these findings with the DAO literature and DAO characteristics to develop a new understanding of the business value of DAOs. Ultimately, the study emphasizes the importance of driving organizational evolution, contributing to the technology management literature, while, asking the following research question:

How do DAOs change the current understanding of business value with regard to open systems?

In summary, the aim of this study is to provide a holistic understanding of DAOs' transformative business potential and to explore the relationship between open business models (OBMs) and DAOs' business value. To achieve this aim, chapter 1 introduces the problem and aim of this research. Chapter 2 provides a theoretical background and identifies the design characteristics of OBMs and open innovation. Chapter 3 provides the methodology of the best-fit framework synthesis and explains the research steps taken. Chapter 4 synthesizes the DAO literature and the open business model characteristics, to create a new framework in chapter 5. Chapter 6 will identify gaps in the literature for further research opportunities, while concluding.

## 2      Background

We understand BMs as business architectures (Teece, 2010; Osterwalder et al., 2005) and use the term, corresponding with Zott and Amit (2010), in the following sense: (1) The BM is a *unit of analysis*; (2) emphasizing a *system-level* outlook, explaining how organizations *do business*; (3) involving *boundary-spanning* interdependent activities undertaken by a focal firm or other entities; (4) in order to *create deliver and capture value*. Berglund & Sandström (2013) point to the definition as *"a system of interdependent activities that transcends the focal firm and spans its boundaries"* (Zott & Amit, 2010, p. 1) while further arguing, that the systemic and boundary-spanning nature of BMs imply that firms act under conditions of interdependence and restricted freedom, since they do not have executive control over their surroundings. In this sense, the authors introduce the term *open systems*, which can be understood as a systemic view of BMs, facilitating interactions between the focal subject or organization and its surrounding environment.

Value creation is the ultimate objective of all corporations (Herskovits et al., 2013), and Climent et al. (2021) emphasize that the traditional understanding of value creation patterns assumes that the process of value creation is limited to the borders of an organization through the firm's activities. In this context, Redlich & Moritz (2016) describe the evolution of value creation, beginning with traditional methods centered on companies and passive customer roles, advancing to value creation networks for collective efficiency. They emphasize the shift towards interactive and co-creative approaches, where customers become more active participants, resulting in the dissolution of hierarchical structures and the involvement of various actors in the value creation process of a firm, including suppliers, customers, users, and community members. This shift in perspective has given rise to value co-creation as a new paradigm, which views consumers not merely as purchasers, but as knowledgeable users who contribute across all phases of the value creation process of an organization (Grabher & Ibert, 2018). As Chesbrough and Appleyard (2007, p. 5) put it, "shifting the focus from closed structures to the concept of openness requires a reconsideration of the processes that underlie value creation and value capture", and thereby requires examination of the matter at the level of a BM. Value capture, like value creation, is an integral part of every BM, explaining how an organization captures value from its value creation (Osterwalder & Pigneur, 2010). Though, while value creation has gained extensive academic attention since the early 2000s (Raasch & Herstatt, 2011), the exploration of value capture in open systems has lagged (West & Bogers, 2014). However, there is widespread recognition that substantial value can be generated within open organizational structures (Wasko and Faraj, 2005), though viable BMs integrating value capture in open structures remain less clear (Stefan et al., 2022) and the questions revolving around how to appropriate returns in open systems have been described as the 'paradox of openness' (Laursen & Salter, 2014). Hence, Toroslu et al. (2023) underscore the importance of comprehending the distribution of value within open systems to ensure effective value capture mechanisms, being able to avoid failures in open organizational structures. As systems shifted from closed to open, the concept of value evolved, necessitating collaboration among distributed actors in a multi-actor interdependent value exchange process (Chesbrough et al., 2018). These values can represent a sequence of activities in a BM via a systematic approach, known as a value system. Accordingly, these values are categorized into value capture, value creation and delivery, and value proposition (Täuscher & Laudien, 2018).

In summary, organizations seeking to engage in open systems must adapt their BM to adhere to open business principles (Saebi & Foss, 2015). This process of reorganizing BM components is known as business model innovation (BMI) (Foss & Saebi, 2017). In line with Sjödin et al. (2020), this study enhances the understanding of the alignment of value creation and value capture in open innovation processes. We highlight the collaborative nature of open BMI while focusing on DAO characteristics involving customers, users, community members, and organizations. Hereby, we follow recent research endeavors (Chesbrough et al., 2018; Foss & Saebi, 2017; Sjödin et al. 2020) that argue for a stronger emphasis on the external perspective, especially concerning understanding the role of community interactions in shaping BMI and restructuring the organizational value systems.

## 3    Methodology

To gain a deeper understanding of the emerging research field around DAOs' business value, we opted for a systematic review of the literature using a framework synthesis method (Dixon-Woods, 2011). As Shaw et al. (2020) put it, the best-fit framework synthesis uses a deductive approach of mapping data from primary research studies onto a framework constructed of pre-identified themes, concepts, theories or ideas, and later undergoes an inductive, iterative analysis based on the gathered evidence. Therefore, the methodology produces an updated framework that may incorporate a mix of initial and revised themes, while also requiring two parallel and distinct search strategies: One for identifying the models and frameworks to generate an *a priori framework*, and a second search for populating the primary research studies and being able to extract study data (Carroll et al., 2013) which can be seen in *table 1*.

**Table 1.** Research model.

| *Steps 1-7: "best-fit" framework synthesis* | | *Methods* |
|---|---|---|
| 1: Identify the research question | | Review Protocol, PICOs |
| 2a: Search for models and frameworks | 2b: Search for primary studies | BeHEMoTh & Systematic Review |
| 3a: *A priori framework* | 3b: *Data extraction* | Framework & thematic analysis |
| 4: *Code data* from primary studies against the *a priori framework* | | Deductive thematic analysis |
| 5: Create new themes that cannot be coded against the framework | | Inductive thematic analysis |
| 6: Synthesize *new framework* composed of a priori and new themes | | Framework synthesis |
| 7: Discuss and explore findings | | Generalizations, dissemination |

**2a: Search for models and frameworks.** To develop our *a priori framework,* we followed the systematic BeHEMoTh search procedure as advised by Carroll et al. (2013) along with the steps outlined in the methodological framework by Booth and Carroll (2015). After (re)defining our research question, we conducted our first set of search queries in three databases EBSCO, Google Scholar, and Scopus: (in title: („value creation" OR „ value capture" OR „value system" OR „business value" AND „open" )) AND (in abstract: („model" or „framework"). The files were merged and screened for potential duplicates. From 211 initial files, we found 25 duplicates, which left 186 files that we analyzed, initially going through titles and abstracts. After, we conducted a full-text review of 53 records matching our search terms and the research question. This left us with 3 applicable papers with frameworks for open systems. After an extensive forward and backward search, we included an additional position, depicted in our a priori framework in *table 2*.

**Table 2.** *A priori framework.*

| Layer 1: Open business model strategies | | Author(s) |
|---|---|---|
| Efficiency-centric<br>Collaborative | Crowd-based<br>Network-based | Saebi & Foss (2015) |
| Layer 2: Open business model values | | |
| Value creation<br>Value delivery | Value capture<br>Value proposition | Gassmann et al. (2017);<br>Täuscher & Laudien (2018) |
| Layer 3: Affected business transactions | | |
| Content<br>Governance | Structure | Zott & Amit (2010) |

**3a: A priori framework.** Our *a priori* framework is divided into three layers, comprising of different elements which all need to be considered when thinking about structuring an OBM. On the first layer, an OBM requires a focal organization to strategize how to design the inflow and outflow of knowledge to facilitate the participation with externals: *efficiency-centric*, *crowd-based*, *collaborative* or *network-based* strategies can be considered (Saebi & Foss, 2015). The second layer comprises different internal and external stakeholder values that need to be aligned with organizational goals for effective value creation within the OBM. These values involve defining the *value proposition* of the organization, and implementing a method for *value creation*, *value capture* and *value delivery* (Gassmann et al., 2017; Täuscher & Laudien, 2018). For the third layer, we list the three transactional dimensions which are affected by the exploitation of business opportunities. Here, we consider *content* (elemental activities), *structure* (organizational units performing activities), and *governance* (mechanisms for controlling units and activities) (Zott & Amit, 2010).

**2b and 3b: Search for primary studies and extract data.** To identify our primary studies, we conducted a secondary set of search queries, utilizing our developed PICOs and the search terms: (("business value" OR "value capture" OR "value creation") AND (DAO OR DAOs OR "decentralized autonomous organization")) across three databases: EBSCO, Google Scholar, and Scopus. These queries yielded a total of 2550 studies, which underwent title and abstract screening, resulting in a reduction to 202 articles. Subsequently, we merged the files and conducted another round of screening for duplicates, yielding 171 unique files for analysis. Following this initial screening, we performed a full-text review aligned with our search criteria and research question, identifying 28 papers. Additionally, employing a snowball search, uncovering 8 additional studies, leading to the review of 36 studies of which we extracted study data, lastly including 19 academic papers and 4 practical documents.

## 4    *Code data* **against** *a priori framework*

**Layer 1: Open business model strategies.** By integrating DAOs into innovation strategies, organizations can benefit from reduced coordination costs, enhanced transparency, and increased engagement with external stakeholders. DAOs create an environment for fostering co-creation and knowledge exchange (Bellavitis et al., 2023; Mačiulienė & Skaržauskienė, 2021), achieving both cooperation and coordination (Lumineau et al., 2021). As such, DAOs represent a valuable tool for organizations aiming to adopt open innovation strategies and boost community engagement. **Efficiency-centric.** In an efficiency-centered market-based strategy, efficiency is achieved by reducing transaction and coordination costs, and redefining internal R&D systems (Saebi & Foss, 2015). Working well for DAOs, for instance operating as for-profit venture platforms (e.g. *The LAO*), this strategy enables members to collectively decide on funding early-stage blockchain platforms through a voting mechanism, while streamlining transactions and reducing coordination costs through decentralized decision-making (Ding et al., 2023; The LAO, 2021). **Crowd-based.** In a crowd-based open innovation strategy, input from user communities drives ideation, outsourced to the crowd, while incentives, such as rewards, are given to external knowledge providers (Saebi & Foss, 2015). This strategy can be found in non-profit grant DAOs (e.g. *Moloch DAO*) to fund public infrastructure development. It attracts crowd-contributions from developers and donors through open-source proposals, fostering community involvement and innovation (Santana & Albareda, 2022; Soleimani et al., 2019). **Collaborative.** In a collaborative open innovation strategy, users, suppliers, customers, and competitors play pivotal roles as partners, while sharing values through contracts as external knowledge providers (Saebi & Foss, 2015). This strategy can be found in media platforms (e.g. *Bankless DAO),* whereby, various stakeholders from diverse backgrounds work on joint projects, fostering a culture of open collaboration. Hence, the accessibility to external contributors opens the possibility to specialized expertise, enhancing the quality and impact of the collaborative project (Ding et al., 2023; Bankless DAO, n.d.). **Network-based.** A network-based open innovation strategy involves deeply integrating many external partners into the organization's innovation activities, facilitating the co-development through open platforms (Saebi & Foss, 2015). DAO open-source infrastructures (e.g. *Aragon*) use this strategy to engage its community in their innovation efforts, whereby participants collectively contribute to Aragon's development, leveraging their expertise to drive ecosystem innovation (Saurabh et al., 2023; Aragon, 2020).

**Layer 2: Open business model values**. DAO BMs can profoundly shape an organization's value system by introducing decentralized decision-making, empowering community members to collaboratively shape its direction of values, thus fostering collaboration and inclusivity (Purusottama et al., 2022a). **Value creation.** DAOs can create value through decentralized applications (dApps) and services tailored to users' needs (Schlecht et al., 2021). Further, broad community engagement drives value generation for all stakeholders by leveraging collective expertise and shared values (Upadhyay, 2020). **Value capture.** DAOs value capture mechanisms involve tokens and require defined token models, which include token issuance, governance, and revenue possibilities (Chen, 2018). Further, the token can incentivize community engagement and facilitate the organizations' self-operation and self-governance (Santana & Albareda, 2022). **Value delivery.** DAOs deliver value to users enabling multi-actor interaction through a decentralized platform, while facilitating efficient resource allocation and community-driven initiatives (Purusottama et al., 2022a). Additional advantages for value delivery in DAOs are enhanced by facilitating economic redistribution of values and empowering all nodes to participate in the BM activities (Chen & Bellavitis, 2020). **Value proposition.** DAOs enable organizations to streamline decision-making processes, reduce administrative overhead, and initiate trust via blockchain between the users in the multi-agent organization, alongside efficiency gains from decentralized decision-making (Lage et al., 2022). Additionally, organizations with DAO characteristics offer a new spectrum of possibilities for co-creation and open innovation, enabling superior levels of transparency, security, and fairness compared to traditional BMs (Mačiulienė & Skaržauskienė, 2021).

**Layer 3: Affected business transactions.** DAO BMs decrease transaction costs, establish distributed trust, offer transparency, and facilitate fast, secure, and immutable exchanges (Lage et al., 2022). **Content.** DAOs facilitate the creation of novel transaction types and functionalities through decentralized applications and smart contracts, unlocking new potentials for the exchange and collaboration of digital values (Schlecht et al., 2021), hereby, the blockchain's transparent trail of transactions enables effective auditing and logging, fostering the development of shared BMs (Upadhyay, 2020). **Structure.** DAOs eliminate the power of intermediaries, resulting in cost reduction (Bellavitis et al., 2023), while redefining transactional structures enabling more dynamic networks, and replacing traditional and linear BMs, allowing hypoconnectivity (Taherdoost & Madanchian, 2023). **Governance.** DAOs empower community members to collectively determine transactional rules, fostering more inclusive governance (Chen & Bellavitis, 2020). Organizations leveraging decentralized governance models ensure stakeholder alignment of interests and values, emphasizing community engagement in shaping governance (Saurabh et al., 2023).

## 5        New themes & synthesis of new framework

While OBMs blur the lines between external and internal value creation (Grabher & Ibert, 2018), DAOs advance the concepts of OBMs by including all involved parties as stakeholders in value capture mechanisms and reshaping organizational boundaries (Mačiulienė & Skaržauskienė, 2021). The *DAO business framework* elements outlined in *figure 1* align with open concepts predating blockchain, with blockchain further enhancing trust and transparency in *transactions,* fostering collaboration without intermediaries (Upadhyay, 2020). DAOs facilitate governance through community decision-making (Purusottama et al., 2022b), contrasting with traditional *open strategies* where firms solely interact externally for knowledge (Saebi & Foss, 2015).

Exchanging values in blockchain environments create new *value system mechanisms* that attempt to bypass the "paradox of openness" (Laursen & Salter, 2014). In DAOs this new medium of exchange is called *a token*. Tokens hold the possibility for community members to partake in an organization's transactional activities, while allowing firms to build more robust value systems with greater network effects (Abdollahi et al., 2023). Since tokens influence, user engagement, value management, and are critical for organizational adoption and growth (Upadhyay, 2020), careful consideration needs to be given on defining token systems, token types, incentive systems, and supply and demand dynamics (Weking et al., 2020). Hence, tokens are vital for DAO BMIs and are the fourth element in our DAO business framework.

**Figure 1.** *DAO business framework.*

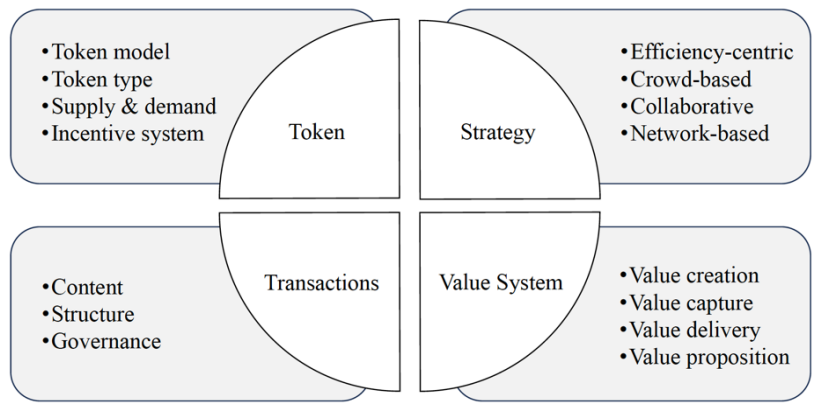

## 6        Conclusion

Given the novelty of the research topic, there are several opportunities for the further exploratio*n:* firstly, there's a need to delve into the diverse factors of stakeholders in multi-agent DAO BMs to enhance participation and collaboration, aligning incentive structures with value management systems to prevent failures in open systems. Secondly, categorizing DAO business model patterns through case analysis can reveal their unique traits, identifying factors influencing success or failure. Lastly, while the conceptual framework offers insights into understanding DAOs' business value, future research should explore its utility in real-world organizational contexts.

In answering the research question on how DAOs change the current understanding of business value with regard to open systems, this research responds to the growing need for research on DAOs' applicability and scalability. The contribution of this study is twofold: First, the preliminary *DAO business framework* contributes to better understanding of DAOs' transformative business potential and provides organizations guidance on altering organizational power dynamics helping to gain access to communities and collective wisdom, which in turn can lead to an increase of business value. Second, the systematic review provides a foundation for researchers to tackle challenges in the evolving landscape of decentralized organizational activities, particularly in understanding how communities influence BMI and reshape organizational business frameworks.